# Universal Finite-size Scaling Functions with Exact Non-universal Metric Factors

Ming-Chya Wu,* Chin-Kun Hu,† and N. Sh. Izmailian

*Institute of Physics, Academia Sinica, Nankang, Taipei 11529, Taiwan*

(Dated: March 10, 2003)

## Abstract

Using exact partition functions and finite-size corrections for the Ising model on finite square, plane triangular, and honeycomb lattices, we obtain universal finite-size scaling functions for the specific heat, the internal energy, and the free energy of the Ising model on these lattices with exact non-universal metric factors.

PACS numbers: 05.50.+q, 75.10.-b

Finite-size scaling has been of interest to scientists working on a variety of critical systems or problems, including superfluid, spin models, percolation models, lattice gauge models, spin glass, metastability, etc [1–6]. Universal finite-size scaling and finite-size corrections in finite critical systems have attracted much attention in recent decades [2–11]. Using renormalization group (RG) arguments, in 1984 Privman and Fisher first proposed the concepts of universal finite-size scaling functions (UFSSF's) and non-universal metric factors [2]. Using Monte Carlo methods [12] and choosing aspect ratios of the square (sq), plane triangular (pt), and honeycomb (hc) lattices so that they have the relative proportions $1 : \sqrt{3}/2 : \sqrt{3}$ [13], Hu, Okabe, et al. found UFSSF's for the percolation and the Ising models on two-dimensional lattices [5, 6]. Since these studies were based on numerical simulations, there are always some numerical uncertainty in the obtained results. Here we use the exact partition functions of the Ising model on finite sq, pt and hc lattices with periodic-aperiodic boundary conditions [14, 15] and an exact expansion method [10] to obtain exact finite-size corrections of the free energy $f^B$, the internal energy $E^B$, and the specific heat $C^B$ of the critical Ising model on these lattices with periodic-periodic (pp), periodic-antiperiodic (pa), antiperiodic-periodic (ap) and antiperiodic-antiperiodic (aa) boundary conditions, where $B$ denotes boundary conditions. Using these coefficients and extend a method [16] to subtract leading singular term from the free energy, we subtract leading singular terms from $C^B$, $E^B$, and $f^B$ and find that the corresponding residuals $\Delta^B$, $\Gamma^B$, and $W^B$ (see Eqs.(7)-(9) below), have very good finite-size scaling behavior. Choosing aspect ratios of the sq, pt, and hc lattices so that they have the relative proportions $1 : \sqrt{3}/2 : \sqrt{3}$ [13] and calculating exact non-universal metric factors from exact partition functions and finite-size corrections for the Ising model, we find that $\Delta^B$, $\Gamma^B$, and $W^B$ of the sq, pt, and hc lattices have very nice universal finite-size scaling behavior.

The partition function of the Ising model is given by

$$Z = \sum_{\{\sigma\}} \exp[\beta J \sum_{\langle ij \rangle} \sigma_i \sigma_j], \qquad (1)$$

where $J$ is the coupling constant of the Ising spin, $\beta = 1/(k_B T)$ with $k_B$ being the Boltzmann constant and $T$ being the absolute temperature, $\sigma_i = \pm 1$ is the Ising spin at site $i$, the first sum is over all spin states, and the second sum is over the nearest-neighbor (NN) pairs $< i, j >$ of the Ising spins. In this Letter, the geometric structures of the sq, pt and hc lattices follow those in [13]. The exact partition functions for $L_1 \times L_2$ sq, pt and hc lattices

TABLE I: Sign factors $\varepsilon_i^B$ in (2) for various BC's.

| B.C. | $\varepsilon_1^B$ | $\varepsilon_2^B$ | $\varepsilon_3^B$ | $\varepsilon_4^B$ |
|---|---|---|---|---|
| pp | + | + | + | + |
| pa | + | + | − | − |
| ap | + | − | + | − |
| aa | − | + | + | − |

TABLE II: Expressions for $g$, $A_0$, $A_1$, $A_2$ in (3) and $\beta_c J$ for various lattices, $t = \tanh(\beta J)$.

| lattice | $g$ | $A_0$ | $A_1$ | $A_2$ | $\beta_c J$ |
|---|---|---|---|---|---|
| sq | 1 | $\left(1+t^2\right)^2$ | $2t\left(1-t^2\right)$ | 0 | $\frac{1}{2}\ln\left(1+\sqrt{2}\right)$ |
| pt | 1 | $\left(1+t^2\right)^3 + 8t^3$ | $2t\left(1-t^2\right)^2$ | $2t\left(1-t^2\right)^2$ | $\frac{1}{2}\ln\sqrt{3}$ |
| hc | $\frac{1}{2}$ | $1+3t^4$ | $2t^2\left(1-t^2\right)$ | $2t^2\left(1-t^2\right)$ | $\frac{1}{2}\ln\left(2+\sqrt{3}\right)$ |

for periodic-aperiodic boundary conditions have been found to be [15]

$$Z^B = \frac{1}{2} 2^{L_1 L_2} \left[\cosh(\beta J)\right]^{zL_1 L_2/2} \sum_{i=1}^{4} \varepsilon_i^B \Omega_i, \qquad (2)$$

where $z$ is the coordination number of the lattice and $\varepsilon_i^B$ shown in Table I are sign factors specified by boundary conditions (BC's), and $\Omega_1 = \Omega_{\frac{1}{2}\frac{1}{2}}$, $\Omega_2 = \Omega_{\frac{1}{2}0}$, $\Omega_3 = \Omega_{0\frac{1}{2}}$, $\Omega_4 = -\text{sgn}\left(\frac{\theta-\theta_c}{\theta_c}\right)\Omega_{00}$, with

$$\Omega_{\mu\nu} = \prod_{p=0}^{gL_1-1} \prod_{q=0}^{L_2-1} \left\{ A_0 - A_1 \left[\cos\frac{2\pi(p+\mu)}{gL_1} + \cos\frac{2\pi(q+\nu)}{L_2}\right] - A_2 \cos\left(\frac{2\pi(p+\mu)}{gL_1} - \frac{2\pi(q+\nu)}{L_2}\right)\right\}^{1/2}. \qquad (3)$$

The values of $g$, the explicit forms of $A_0$, $A_1$ and $A_2$, and the critical values of $\beta J$, $\beta_c J$, are listed in Table II.

To calculate the specific heat, we write $Z^B$ as

$$Z^B = \frac{1}{2}\left[2\sinh(2\eta)\right]^{L_1 L_2/2} \sum_{i=1}^{4} \varepsilon_i^B Z_i(\tau, R, L), \qquad (4)$$

where $\eta = \beta J$, $Z_i(\tau, R, L) = \Omega_i/(A_1/2)^{gL_1 L_2/2}$, $\tau = D_2 L\epsilon$, $D_2$ is a non-universal metric factor [17], $L = (L_1 L_2)^{1/2}$ is the characteristic size of the lattice, $\epsilon = (T - T_c)/T_c$ is the reduced temperature and $R = L_1/L_2$ is the aspect ratio. Then, the specific heat

TABLE III: Expressions for $c_{0,0}$, $c_{0,1}$, $c_{0,2}$, $c_1$, $a_{0,0}$, $a_0$, $a_1$, $q$, and $E_{0,0}$; $\gamma_E (= 0.5772156649\ldots)$ is the Euler constant.

| lattice | $c_{0,0}$ | $c_{0,1}$ | $c_{0,2}$ | $c_1$ | $a_{0,0}$ | $a_0$ | $a_1$ | $q$ | $E_{0,0}$ |
|---|---|---|---|---|---|---|---|---|---|
| sq | $\frac{8}{\pi}$ | $\frac{8}{\pi}\left(\ln\frac{4\sqrt{2}}{\pi} + \gamma_E - \frac{\pi}{4}\right)$ | 4 | $2\sqrt{2}$ | $\frac{24\sqrt{2}}{\pi}$ | 16 | 6 | $e^{-\pi R}$ | $-\sqrt{2}$ |
| pt | $\frac{12\sqrt{3}}{\pi}$ | $\frac{12\sqrt{3}}{\pi}\left(\ln\frac{4\sqrt{3}}{\pi} + \gamma_E - \frac{\sqrt{3}\pi}{6}\right)$ | 9 | 6 | $\frac{72\sqrt{3}}{\pi}$ | 54 | $4\sqrt{3}$ | $e^{-\pi(\sqrt{3}-i)R/2}$ | $-2$ |
| hc | $\frac{2\sqrt{3}}{\pi}$ | $\frac{2\sqrt{3}}{\pi}\left(\ln\frac{4\sqrt{3}}{\pi} + \gamma_E - \frac{\sqrt{3}\pi}{9}\right)$ | $\frac{3}{4}$ | $\frac{1}{\sqrt{2}}$ | $\frac{6\sqrt{2}}{\pi}$ | $\frac{3\sqrt{3}}{4}$ | $8\sqrt{3}$ | $e^{-\pi(\sqrt{3}-i)R/4}$ | $-\frac{2}{\sqrt{3}}$ |

$C^B/k_B \equiv (\eta^2/L^2)\left(\partial^2 \ln Z^B/\partial \eta^2\right)$ near $\tau = 0$ can be written as

$$\frac{C^B}{k_B} = \left[C_{0,0}\ln L + C_0^B(R) + \frac{C_1^B(R)}{L} + O\left(\frac{1}{L^2}\right)\right]$$
$$+ \frac{1}{D_2}\left[A_0^B(R) + \frac{A_{0,0}(R)}{L}\ln L + O\left(\frac{1}{L}\right)\right]\tau + O(\tau^2), \quad (5)$$

Here $C_{0,0} = c_{0,0}\eta_c^2$, $C_0^B(R) = \eta_c^2[c_{0,1} - 2c_{0,0}f_1 - c_{0,2}Rf_2^2 - \frac{1}{2}c_{0,0}\ln R]$, $C_1^B(R) = -c_1\sqrt{R}\eta_c^2 f_2$, $A_{0,0}(R) = a_{0,0}\sqrt{R}\eta_c^3 - 2c_{0,0}\eta_c^2$, $A_0^B(R) = -\frac{a_0\sqrt{R}\eta_c^3}{\pi}f_2[\pi R f_2^2 + a_1 f_1 - \frac{1}{3}a_1 f_3]$, where $\eta_c = \beta_c J$, $f_1 = \left(\varepsilon_1^B |\theta_3| \ln |\theta_3| + \varepsilon_3^B |\theta_4| \ln |\theta_4| + \varepsilon_2^B |\theta_2| \ln |\theta_2|\right)/F$, $f_2 = \left(\varepsilon_4^B |\theta_2| |\theta_3| |\theta_4|\right)/F$, $f_3 = \ln(4|\theta_2||\theta_3||\theta_4|)$, $F = \left(\varepsilon_1^B |\theta_3| + \varepsilon_3^B |\theta_4| + \varepsilon_2^B |\theta_2|\right)$, with $\theta_i = \theta_i(0, q)$ the Elliptic theta functions of modulus $q$. Expressions for $c_{0,0}$, $c_{0,1}$, $c_{0,2}$, $c_1$, $a_{0,0}$, $a_0$, $a_1$, and $q$ are listed in Table III.

Based on [16], we write the free energy $f^B \equiv (1/L^2) \ln Z^B$ near $\epsilon = 0$ as

$$f^B \approx F_0^B + F_1^B \epsilon - \frac{1}{2}\left[C_0^B(R) + C_{0,0}\ln L\right]\epsilon^2 - (D_2 L)^{-2} W^B(\tau, R, L), \quad (6)$$

with scaling function $W^B$. Here $F_0^B = f_{0,0} + f_{0,1}/L^2 + O(L^{-4})$ and $F_1^B = \beta_c E_{0,0} + \beta_c E_{0,1}/L + O(L^{-2})$, where $f_{0,0} = 0.9296954\ldots$ for sq, $0.8795854\ldots$ for pt, and $1.0250591\ldots$ for hc lattices, $f_{0,1}^B = \ln F - \frac{1}{3}\ln f_3$, $E_{0,0}$ for the sq, pt and hc lattices are listed in Table III, and $E_{0,1}^B = -\sqrt{c_{0,2}R}f_2$.

To study finite-size scaling, we define the scaling function for the specific heat, $\Delta^B$, as

$$\Delta^B(\tau, R, L) = \frac{\partial^2 W^B(\tau, R, L)}{\partial \tau^2} = \frac{c^B}{k_B} - \left[C_0^B(R) + C_{0,0}\ln L\right]$$
$$= \frac{1}{D_2}\left[A_0^B(R) + \frac{A_{0,0}(R)}{L}\ln L + O\left(\frac{1}{L}\right)\right]\tau + O(\tau^2) + \frac{C_1^B(R)}{L} + O\left(\frac{1}{L^2}\right). \quad (7)$$

According to the definition of internal energy $E^B = -(1/L^2)\left(\partial \ln Z^B\right)/(\partial \eta)$, we define the scaling function for the internal energy, $\Gamma^B$, as

$$\Gamma^B(\tau, R, L) = \frac{\partial W^B(\tau, R, L)}{\partial \tau} = -D_2 L \beta_c \left\{ E^B - \left( E_{0,0} + \frac{E^B_{0,1}}{L} \right) + \frac{1}{\beta_c} \left[ C^B_0(R) + C_{0,0} \ln L \right] \epsilon \right\}$$
$$\simeq \Delta^B \tau + O(\tau^2). \tag{8}$$

Similarly, the scaling function for the free energy, $W^B$, is

$$W^B(\tau, R, L)$$
$$= -(D_2 L)^2 \left\{ \frac{\beta_c}{\beta} \left[ f^B - \left( f_{0,0} + \frac{f^B_{0,1}}{L^2} \right) \right] + \beta_c \left( E_{0,0} + \frac{E_{0,1}}{L} \right) \epsilon + \frac{1}{2} \left[ C^B_0(R) + C_{0,0} \ln L \right] \epsilon^2 \right\}$$
$$\simeq \frac{1}{2} \Delta^B \tau^2 + O(\tau^3). \tag{9}$$

In the scaling function $\Delta^B$, $C^B_1(R)$ is the leading term in finite-size corrections, and can be used to interpret the finite-size effects of the lattice systems. From this aspect, the finite-size effect for the pp boundary condition is the smallest in comparison with the pa, ap and aa boundaries. The behaviors of $\Delta^B(\tau, R = 1, L)$, $\Gamma^B(\tau, R = 1, L)$ and $W^B(\tau, R = 1, L)$ as a function of $\tau$ with $D_2 = 1$ for the sq, pt and hc lattices with periodic boundary conditions are shown in Figs.1(a), (b) and (c), respectively, which show that these quantities have very nice finite-size scaling behavior. We note that, for isotropic coupling, $\Delta^{\text{pa}}_{\text{sq}}(\tau, R = 1, L) = \Delta^{\text{ap}}_{\text{sq}}(\tau, R = 1, L)$ for the sq lattice, and $\Delta^{\text{pa}}_{\text{pt}}(\tau, R = 1, L) = \Delta^{\text{ap}}_{\text{pt}}(\tau, R = 1, L) = \Delta^{\text{aa}}_{\text{pt}}(\tau, R = 1, L)$ for the pt lattice, due to their symmetric structures. However, $g \neq 1$ for the hc lattice violates the symmetry by exchanging $L_1$ and $L_2$ when $R = 1$, and $\Delta^{\text{pa}}_{\text{pt}}(\tau, R = 1, L) \neq \Delta^{\text{ap}}_{\text{pt}}(\tau, R = 1, L) \neq \Delta^{\text{aa}}_{\text{pt}}(\tau, R = 1, L)$.

To study the UFSSF's of $\Delta^B(\tau, R, L)$, we take aspect ratios of the sq, pt and hc lattices to have the relative proportions $1 : \sqrt{3}/2 : \sqrt{3}$. Equation (7) implies that for large lattices, the slope of $\Delta^B$ as a function of $\tau$ at $\tau = 0$ is determined by $A^B_0$. We can further multiply the scaling function by another non-universal metric factor $D_1$ to obtain

$$D_1 \Delta^B(\tau, R, L) = \frac{D_1}{D_2} \left[ A^B_0(R) + \frac{A_{0,0}(R)}{L} \ln L + O\left(\frac{1}{L}\right) \right] \tau + O(\tau^2) + O\left(\frac{1}{L}\right). \tag{10}$$

We can take values of $D_1/D_2$ so that the leading terms in the right hand sides of (10) have the same slope for sq, pt, and hc lattices. Accordingly, we take the values of $D_1/D_2$ as $(D_1/D_2)_{\text{sq}} = 1$ for the sq lattice, $(D_1/D_2)_{\text{pt}} = A^B_{0,\text{sq}}(R_{\text{sq}})/A^B_{0,\text{pt}}(R_{\text{pt}})$ for the pt lattice and $(D_1/D_2)_{\text{hc}} = A^B_{0,\text{sq}}(R_{\text{sq}})/A^B_{0,\text{hc}}(R_{\text{hc}})$ for the hc lattice. To explore the behaviors of the amplitude $A_0^B$ as a function of the aspect ratio $R$, we consider $A_0^B(D_3R)$ with scale factors $D_3 = 1, \sqrt{3}/2, \sqrt{3}$ for the sq, pt and hc lattices, respectively. The sign of the $A_0^B(D_3R)$, in general, gives the information about the location ($\tau_{\max}$) of the maximum of the specific heat. When $A_0^B(D_3R)$ is positive, $\tau_{\max} > \tau_c = 0$, and when $A_0^B(D_3R)$ is negative, $\tau_{\max} < \tau_c$. For the pp and pa boundary conditions, the behaviors of $A_0^B(D_3R)$ for the sq, pt and hc lattices are roughly in a similar feature in the region of $0 < R \lesssim 1$. However, for the ap and aa boundary conditions, the curves of $A_0^B(D_3R)$ for sq lattice are quite different from those for pt and hc lattices.

To show examples of $(D_1/D_2)$ and UFSSF's, we note that for $R = 1$, $(D_1/D_2)_{\text{pt}} = 0.9688\ldots$(pp), $1.0297\ldots$(pa), and $(D_1/D_2)_{\text{hc}} = 1.0332\ldots$(pp), $1.0981\ldots$(pa). For $R = 1/2$ and $1/3$, we have $(D_1/D_2)_{\text{pt}} = 0.9864\ldots$(pp), $0.9864\ldots$(pa), and $(D_1/D_2)_{\text{hc}} = 1.0520\ldots$(pp), $1.0520\ldots$(pa). We find that the values of $D_1/D_2$ are roughly $0.9864\ldots$, $1.0520\ldots$ for the pt and hc lattices in the range, $0.05 \lesssim R \lesssim 0.75$ and the values are the same for pp and pa boundary conditions. Using these values and calculating $D_1$ based on exact specific heat for such finite lattices, we plot the $D_1\Delta^B(\tau, D_3R, L)$ for the sq, pt and hc lattices in Figs.2(a), (b) and (c), which show that the residual specific heats have very nice universal finite-size scaling behavior. Based on $\Delta^B$, $\Gamma^B$ and $W^B$ in Eqs.(7)-(9), we can further multiply the scaling functions of the internal energy and the free energy by the same non-universal metric factor $D_1$, and obtain the universal scaling functions for $D_1\Gamma^B(\tau, D_3R, L)$ and $D_1W^B(\tau, D_3R, L)$, which are plotted in Figs.3(a), (b) and (c). These figures also show very nice universal finite-size scaling behavior.

Although the results of this Letter are based on analytic expressions for the physical quantities of the Ising model, our formulations can be extended to numerical or experimental studies of critical finite systems [18]. For example, the coefficients $C_0^B(R)$ and $C_{0,0}$ of Eq. (7) can be evaluated by using extrapolation techniques to analyze simulation or experimental data, then one can define the scaling function of the specific heat similar to that in Eq.(7). As a result, the UFSSF for the specific heat can be obtained.

We thank H. W. J. Bloete for a critical reading. This work was supported in part by the National Science Council of the Republic of China (Taiwan) under Grant No. NSC 91-2112-M-001-056.

* Electronic address: mcwu@phys.sinica.edu.tw

† Electronic address: huck@phys.sinica.edu.tw

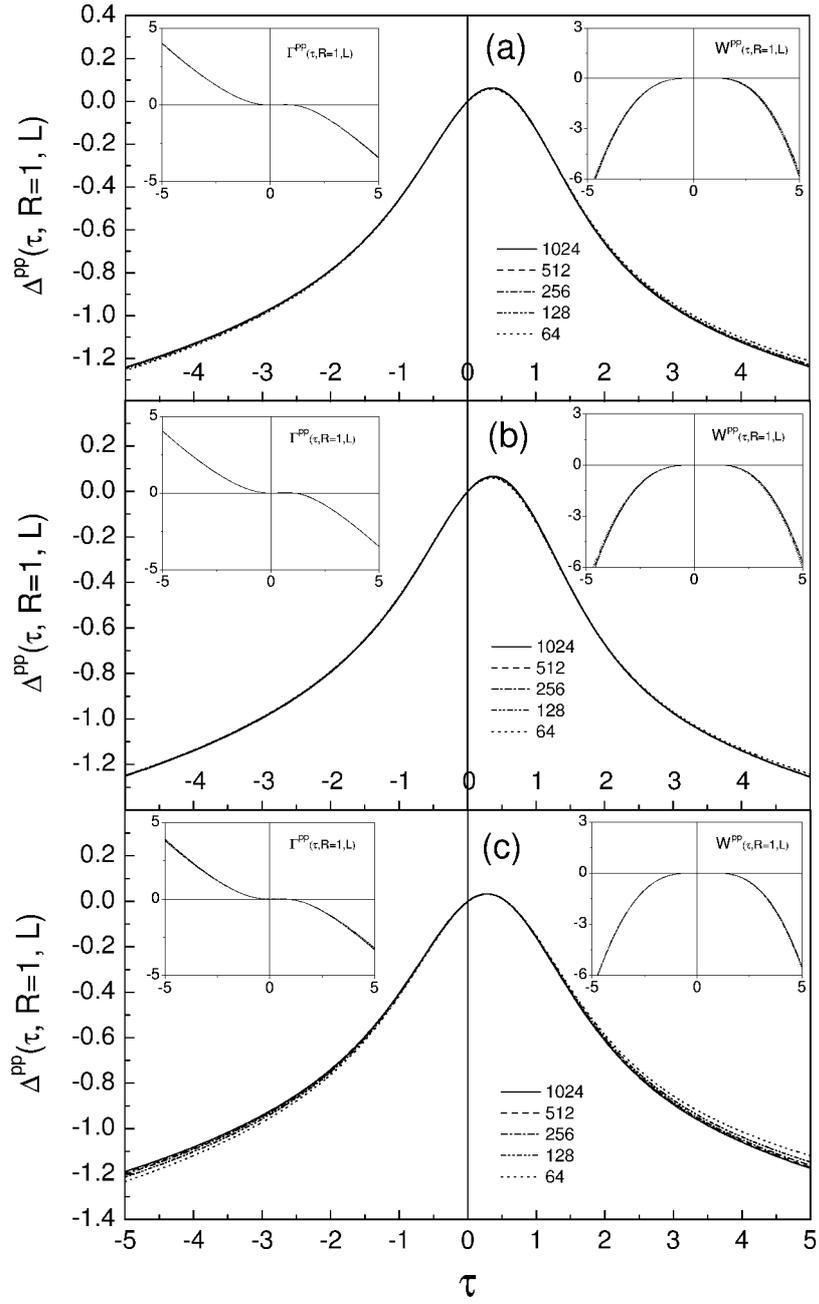

FIG. 1: The scaling functions $\Delta^{pp}(\tau, R=1, L)$, $\Gamma^{pp}(\tau, R=1, L)$, and $W^{pp}(\tau, R=1, L)$ as a function of $\tau$ with $D_2 = 1$ for (a) sq (b) pt, (c) hc lattices under periodic-periodic (pp) boundary condition.

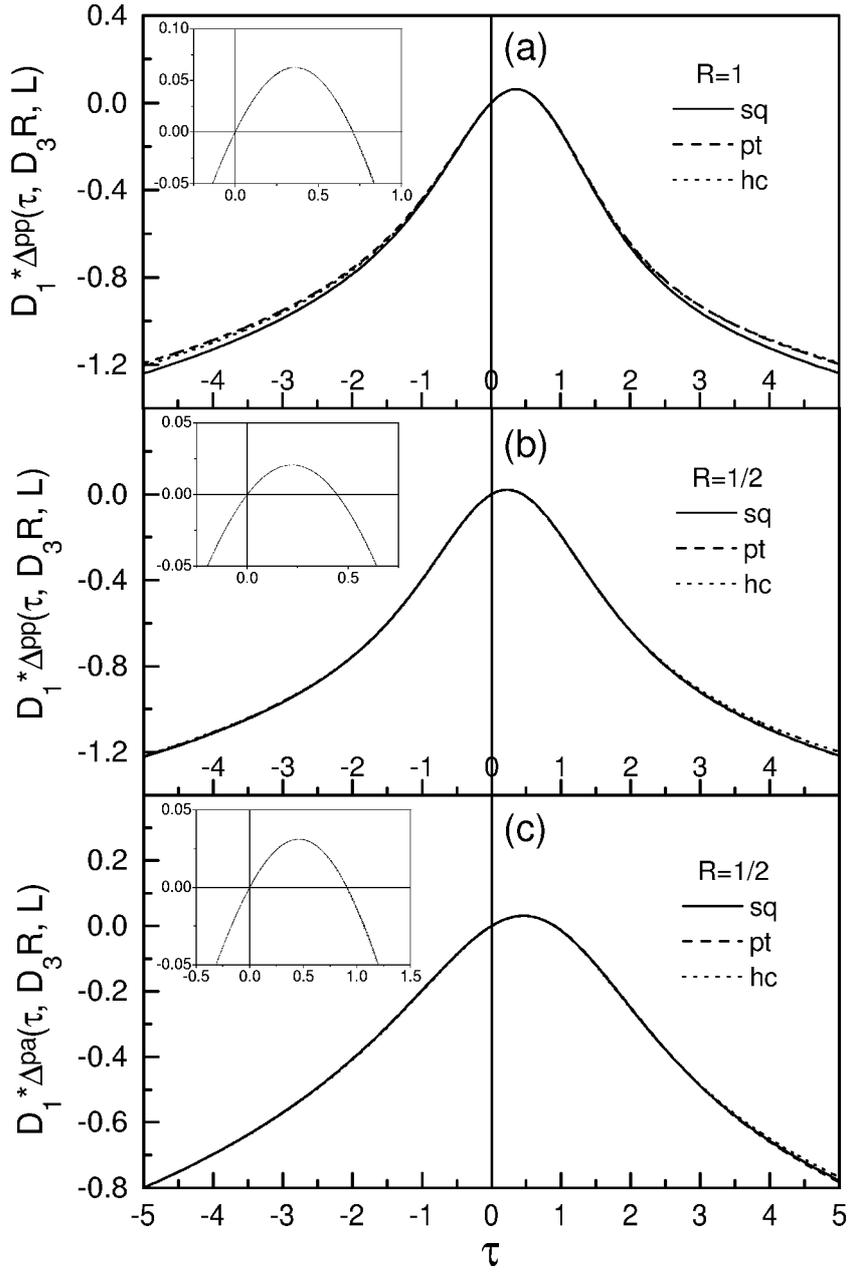

FIG. 2: The scaling function $D_1 \Delta^B (\tau, D_3 R, L)$ as a function of $\tau$ for sq, pt and hc lattices with aspect ratios $R = 1$ and $D_3 = 1, \sqrt{3}/2, \sqrt{3}$, for sq, pt, and hc lattices, respectively, under pp boundary condition and $L_1 = 1024$, $D_1^{pt} = 0.957(\text{pp})$, $D_1^{hc} = 0.9896\ldots(\text{pp})$; and with aspect ratios $R = 1/2$ under (b) pp (c) pa boundary conditions and $L_1 = 768$, $D_1^{pt} = 0.9898\ldots(\text{pp})$, $0.9997\ldots(\text{pa})$, and $D_1^{hc} = 1.018\ldots(\text{pp})$, $1.0167\ldots(\text{pa})$. The insets show curves near $\tau = 0$ in more details.

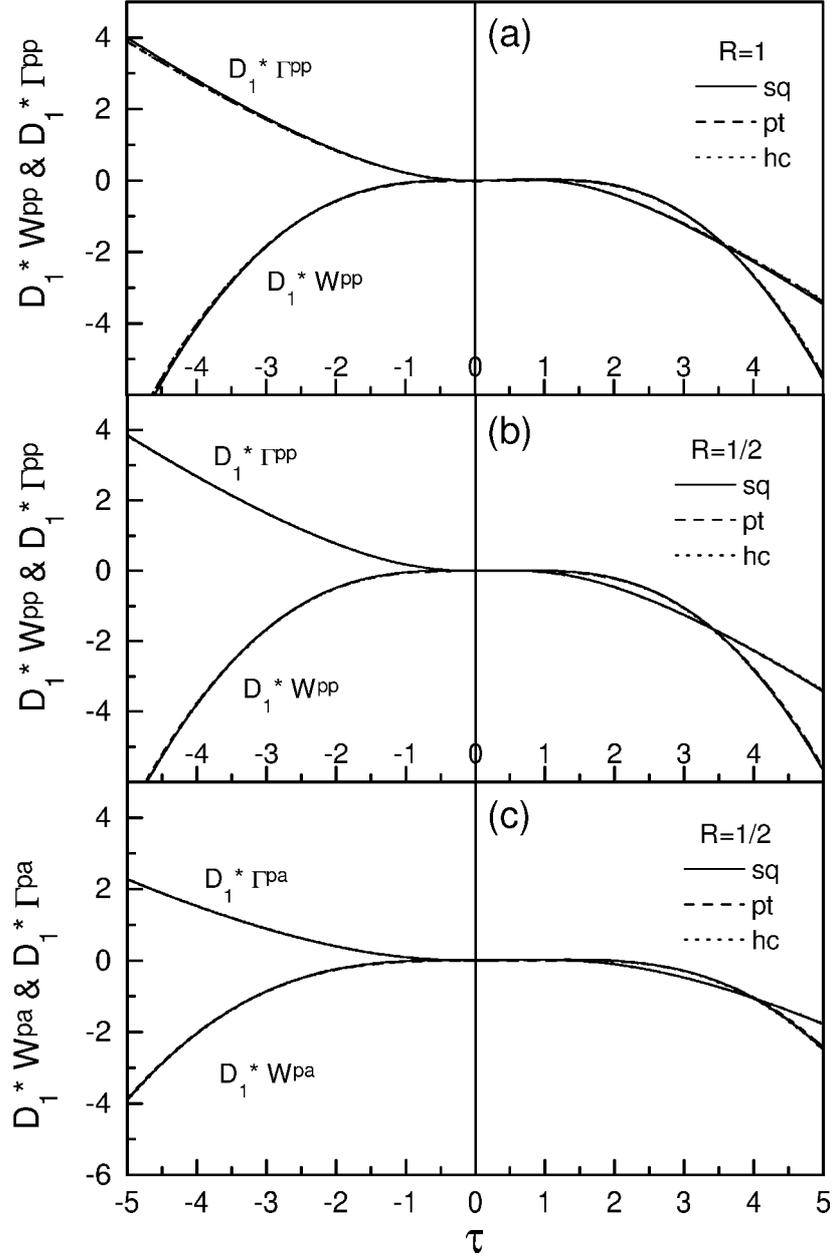

FIG. 3: The scaling functions $D_1 W^B (\tau, D_3 R, L)$ and $D_1 \Gamma^B (\tau, D_3 R, L)$ as a function of $\tau$ for sq, pt and hc lattices with aspect ratios $R = 1$ and $D_3 = 1, \sqrt{3}/2, \sqrt{3}$, for sq, pt, and hc lattices, respectively, under pp boundary condition and $L_1 = 1024$, $D_1^{\text{pt}} = 0.957\ldots(\text{pp})$, $D_1^{\text{hc}} = 0.9896\ldots(\text{pp})$; and with aspect ratios $R = 1/2$ under (b) pp (c) pa boundary conditions and $L_1 = 768$, $D_1^{\text{pt}} = 0.9898\ldots(\text{pp})$, $0.9997\ldots(\text{pa})$, and $D_1^{\text{hc}} = 1.018\ldots(\text{pp})$, $1.0167\ldots(\text{pa})$.